# Study on Graphene based Next Generation Flexible Photodetector for Optical Communication

K. Majumder[1], D. Barshilia[2] and S. Majee[2]

We report on the efficient photodetection (PD) properties of graphene based p-i-n photodetector, where all the three layers are either single or multilayer graphene sheets. We report the bandwidth and responsivity performance of the device. This simple structure paves the way for the next generation flexible wireless communication systems. A theoretical model is used to study the carrier distribution and current in a graphene based p-i-n photodetector system.

*Introduction:* Photodetectors play key role in high performance of optoelectronic and photonic systems. The next generation flexible and wearable communication systems require efficient photodetector which is compatible with the flexible fabrication process. Although, in recent years, graphene based photodetectors have been successfully reported,[1-4] they are mostly fabricated on rigid substrates. Graphene is a zero band gap hexagonal honeycomb carbon atomic layer,[5-11] which allows absorption in broad range of wavelengths. Along with other unique properties of graphene, this wonder material is extremely flexible and it has huge potential to be used in flexible device fabrication. Thus, though there are vast applications of photodetector mainly in the nano-metric dimensions with unique features, several areas require immediate attention to optimize the engineered properties of such devices. Implementing a useful model for graphene based photodetector is still under research. In this letter, we describe a model for graphene based p-i-n photodetector based on simple concepts, where all the layers of the device is made of graphene sheets allowing its applicability in flexible electronics applications.

*Model*: The model is based on the previously reported study on photodetector devices.[12-16] We consider the graphene based p-i-n photodetector structure, as shown in **Fig. 1**. Here, the light is incident on the P side. The structure consists of a single layer of $N^+$ graphene layer, an undoped multi-layer graphene with thickness $l$ and finally a single layer graphene $P^+$ layer. The nominal $N^+$ and $P^+$ region doping is taken of the order of $1 \times 10^{12}$ cm$^{-2}$, which is practically reliable value.[17]

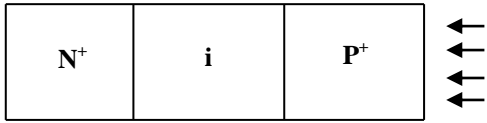

**Fig.1** *Schematic structure of a graphene based p-i-n photodetector.*

In the theoretical model, we have to consider the effect of photogenerated carriers in the intrinsic layer as because the widths of the three regions are comparable to each other. Thus, the current continuity equations in the depletion layer is given by [12]

$$\frac{\partial n(x,t)}{\partial t} - v_n \left\{ \frac{\partial n(x,t)}{\partial x} \right\} = g \quad (1)$$

$$\frac{\partial p(x,t)}{\partial t} + v_p \left\{ \frac{\partial p(x,t)}{\partial x} \right\} = g \quad (2)$$

Where, $g$ is the photo carrier generation rate, $v$ is the velocity, $n$ and $p$ denote the electron and hole, respectively. Incident optical powers, absorption coefficient of graphene at the operating wavelength, reflectivity of the graphene surface are some of the important parameters which control the generation rate of the photo carriers.

Carrier distribution $N(x,j\omega)$ for electrons and $P(x,j\omega)$ for holes in the depletion region (in frequency domain) are calculated by solving the above current continuity equations simultaneously. In general, all the uppercase variables are used to indicate the Laplace transform of the corresponding lowercase variables. The entire depletion region is subdivided into equal energy spacing $(\Delta x)$ for calculation. Along its path of motion, each carrier represents a specific position and energy state in the depletion region of the device. For this reason, each carrier is specifically represented as a function of two indices: one position index *(i)* and one energy index *(j)*. So, we substitute $N(x, j\omega)$ by $N(i,j, j\omega)$ and $P(x, j\omega)$ by $P(i,j,j\omega)$.

To obtain the photo-current density, the carrier distribution $N(i,j, j\omega)$ and $P(i,j,j\omega)$ are multiplied by equal energy spacing $(\Delta x)$ and then summed for all *i* and *j*. The current density $J$ of the device is obtained using Eq. (3)

$$J = \frac{q}{L} \sum_i [\sum_j \{ N(i, j, j\omega) v_n(i) + P(i, j, j\omega) v_p(i) \}] \Delta x(i) \quad (3)$$

Where, *L*, *q* and *v* are the length of the PD, electronic charge and the carrier velocity respectively. We consider here that the carrier velocity is only function of the position. Suffix *n* and *p* is used for electrons and for holes respectively.

*Results and discussions*: The material parameters for the graphene layer have been taken from the literature.[1-4, 17] Using those parameters in this present model, 3-dB bandwidth, frequency response and responsivity of the device have been calculated. The calculated values are justified by the performance of the fabricated graphene based PD devices as described in literature.[1-4]

**Fig. 2** shows the 3-dB bandwidth variation i-layer thickness, where p and n layer consist of single layer graphene sheets (thickness ~ 1 nm). Areas of the PD devices are also varied in this case. Maximum bandwidth (~ 5 GHz) is obtained for the multilayer graphene i-layer (5nm). Effect of device dimension on bandwidth can also be observed from the plot, where smaller sized devices have better bandwidth compare to the larger sized devices.

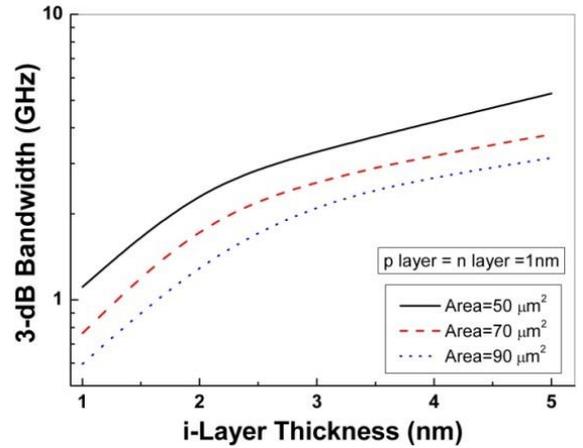

**Fig. 2** *Variation of bandwidth with i-layer thickness.*

**Fig. 3** shows the frequency response, keeping device area fixed at 50 μm2, which gave rise to the maximum 3-dB bandwidth as observed from **Fig. 2**. The i-layer thickness is varied from single layer graphene to few layers of graphene. Again we observe the multilayer graphene i-layer gives rise to better response. This is likely due to the fact that multilayer graphene sheets have poor electrical conductivity as compared to the single layer graphene sheets, giving rise to better performance of the intrinsic layer. Keeping the i-layer thickness as 5 nm and device area as 50 μm$^2$, the responsivity of the system has been investigated. Fig. 4 shows the responsivity of the PD system at 633 nm which is comparable with the reported values in the literature.[4]

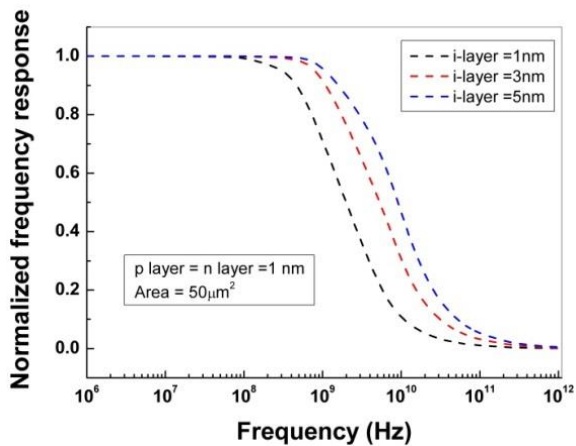

**Fig. 3** *Normalized frequency response with the i-layer thickness.*

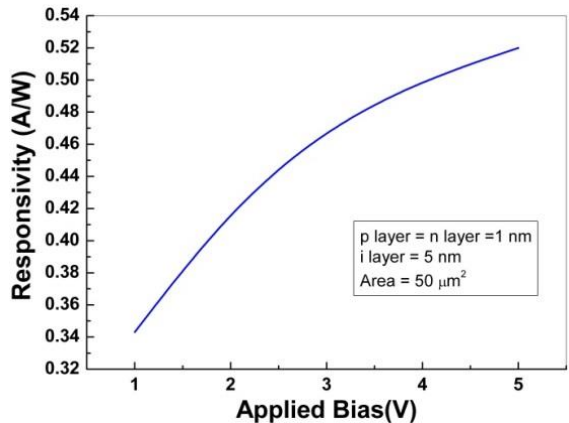

**Fig. 4** *Responsivity of the device at 633nm.*

*Conclusion:* We have designed an effective graphene based flexible PD system whose performance matches well with the experimental values in the literature. The PD system can be effectively used in the next generation communication systems. Further investigations are under progress to fabricate graphene based PD devices on flexible polymer substrates.

*Acknowledgments:* The authors would like to thank Prof. S. Chaudhury, Director CEERI-Pilani and Prof. D. Bhattacharya, Director AOT for their support.

Corresponding author E-mail: *subimal@ceeri.res.in*

K. Majumder: *Academy of Technology, Maulana Abul Kalam Azad University of Technology, G. T. Road, Adisaptagram, Hooghly-712121, West Bengal, India*

D. Barshilia and S. Majee: *CSIR-Central Electronics Engineering Research Institute (CEERI), Pilani 333031, Rajasthan, India*